\documentclass[12pt]{article}
\usepackage{amsmath}
\usepackage{amssymb}
\usepackage{epsfig}
\usepackage{euscript}
\usepackage{fancybox}
\usepackage{color}

\def\mathswitchr#1{\relax\ifmmode{\mathrm{#1}}\else$\mathrm{#1}$\fi}

\newcommand {\pslash}{\hbox{$\not\hbox{\kern-2.3pt $p$}$}}

\usepackage{cite}

\usepackage{epic}

\def\alf1{ {\alpha\over\pi} }

\begin{document}
\begin{titlepage}
\begin{flushright}
{\bf MPI-PhT-2002-08}\\
 {\bf UTHEP-02-0401 }\\
{\bf Mar., 2002}\\
\end{flushright}
 
\begin{center}
{\Large Quantum Corrections to Newton's Law$^{\dagger}$
}
\end{center}

\vspace{2mm}
\begin{center}
{\bf   B.F.L. Ward$^{a,b}$}\\
\vspace{2mm}
{\em $^a$Werner-Heisenberg-Institut, Max-Planck-Institut fuer Physik,
Muenchen, Germany,}\\
{\em $^b$Department of Physics and Astronomy,\\
  The University of Tennessee, Knoxville, Tennessee 37996-1200, USA.}\\
\end{center}


\vspace{5mm}
\begin{center}
{\bf   Abstract}
\end{center}
We present a new approach to quantum gravity starting from
Feynman's formulation for the simplest example, that of a
scalar field as the representative matter. We show that
we extend his treatment to a calculable framework using
resummation techniques already well-tested in other 
problems. Phenomenological consequences for Newton's
law are described.
\vspace{10mm}
\vspace{10mm}
\renewcommand{\baselinestretch}{0.1}
\footnoterule
\noindent
{\footnotesize
\begin{itemize}
\item[${\dagger}$]
Work partly supported 
by the US Department of Energy Contract  DE-FG05-91ER40627
and by NATO Grant PST.CLG.977751.
\end{itemize}
}

\end{titlepage}

\def\Kmax{K_{\rm max}}\def\ieps{{i\epsilon}}\def\rQCD{{\rm QCD}}
\renewcommand{\theequation}{\arabic{equation}}
\font\fortssbx=cmssbx10 scaled \magstep2
\renewcommand\thepage{}
\parskip.1truein\parindent=20pt\pagenumbering{arabic}\par
The law of Newton is the most basic one in physics -- it is already
taught to beginning students at the most elementary level.
Albert Einstein showed that it could be incorporated into his
general theory of relativity as a simple special case
of the solutions of the respective classical field equations.
With the advent of the quantum mechanics of formulations of
Heisenberg  and Schroedinger, one would have thought that
Newton's law would be the first classical law to be 
understood completely from a quantum aspect. This, however, has not happened.
Indeed, even today, we do not have a quantum treatment
of Newton's law that is known to be correct phenomenologically.
In this paper, we propose a possible solution to this problem.
\par

We start from the formulation of Einstein's theory given by Feynman
in Ref.~\cite{f1,f2}. The respective action density is
( in this paper, like Feynman, we ignore matter spin as an inessential
complication~\cite{mlg} )
\begin{equation}
\begin{split}
{\cal L}(x) &= -\frac{1}{2\kappa^2} R \sqrt{-g}
            + \frac{1}{2}\left(g^{\mu\nu}\partial_\mu\varphi\partial_\nu\varphi - m_o^2\varphi^2\right)\sqrt{-g}\\
            &= \quad \frac{1}{2}\left\{ h^{\mu\nu,\lambda}\bar h_{\mu\nu,\lambda} - 2\eta^{\mu\mu'}\eta^{\lambda\lambda'}\bar{h}_{\mu_\lambda,\lambda'}\eta^{\sigma\sigma'}\bar{h}_{\mu'\sigma,\sigma'} \right\}\\
            & \qquad + \frac{1}{2}\left\{\varphi_{,\mu}\varphi^{,\mu}-m_o^2\varphi^2 \right\} -\kappa {h}^{\mu\nu}\left[\overline{\varphi_{,\mu}\varphi_{,\nu}}+\frac{1}{2}m_o^2\varphi^2\eta_{\mu\nu}\right]\\
            & \quad - \kappa^2 \left[ \frac{1}{2}h_{\lambda\rho}\bar{h}^{\rho\lambda}\left( \varphi_{,\mu}\varphi^{,\mu} - m_o^2\varphi^2 \right) - 2\eta_{\rho\rho'}h^{\mu\rho}\bar{h}^{\rho'\nu}\varphi_{,\mu}\varphi_{,\nu}\right] + \cdots \\
\end{split}
\label{eq1}
\end{equation}
Here, $\varphi(x)$ is our representative scalar field for matter,
$\varphi(x)_{,\mu}\equiv \partial_\mu\varphi(x)$,
and $g_{\mu\nu}(x)=\eta_{\mu\nu}+2\kappa h_{\mu\nu}(x)$ is the
metric of space-time where we follow Feynman and expand about Minkowski space
so that $\eta_{\mu\nu}=diag\{1,-1,-1,-1\}$. $R$ is the curvature scalar.
Following Feynman, we have introduced the notation
$\bar y_{\mu\nu}\equiv \frac{1}{2}\left(y_{\mu\nu}+y_{\nu\mu}-\eta_{\mu\nu}{y_\rho}^\rho\right)$ for any tensor $y_{\mu\nu}$\footnote{Our conventions for raising and lowering indices in the 
second line of (\ref{eq1}) are the same as those
in Ref.~\cite{f2}.}. 
Thus, $m_o$ is the bare mass of our matter and we set the small
tentatively observed~\cite{cosm1} value of the cosmological constant
to zero so that our quantum graviton has zero rest mass.
Here, our normalizations are such that $\kappa=\sqrt{8\pi G_N}$
where $G_N$ is Newton's constant.
The Feynman rules for (\ref{eq1}) have been essentially worked out by 
Feynman~\cite{f1,f2}, including the rule for the famous
Feynman-Faddeev-Popov~\cite{f1,ffp1} ghost contribution that must be added to
it to achieve a unitary theory with the fixing of the gauge
( we use the gauge of Feynman in Ref.~\cite{f1}, 
$\partial^\mu \bar h_{\nu\mu}=0$ ), 
so we do not repeat this 
material here. We go instead directly to the treatment of the
apparently uncontrolled UV divergences associated with (\ref{eq1}).\par

\begin{figure}
\begin{center}
\epsfig{file=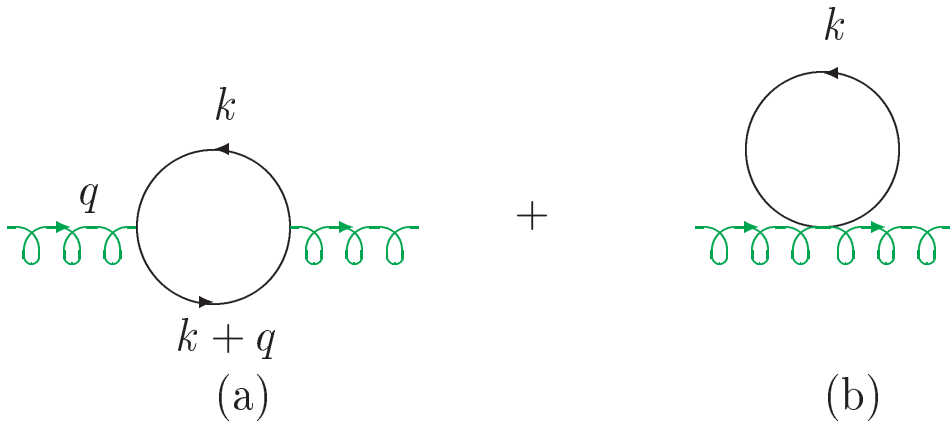,width=140mm}
\end{center}
\caption{\baselineskip=7mm     The scalar one-loop contribution to the
graviton propagator. $q$ is the 4-momentum of the graviton.}
\label{fig1}
\end{figure}

To illustrate our approach, let us study the possible one-loop corrections to
Newton's law that would follow from the matter in (\ref{eq1}) assuming that
our representative matter is really part of a multiplet of fields
at a very high scale ( $M_{GUT}=10^{16}$GeV ) compared to the known SM particles so that it is sufficient to calculate the effects of the diagrams
in Fig.~\ref{fig1}1 on the graviton 
propagator to see the first quantum loop effect.
\par

To this end, we stress the following. The naive power counting of
the graphs give their degree of divergence as +4 and we expect that even
with the gauge invariance there will still remain at least a 0
degree of divergence and that, in higher orders, this remaining
divergence degree gets larger and larger. Indeed, for example, for
Fig. 1a, we get the result 
\begin{equation}
i\Sigma(q)^{1a}_{\bar\mu\bar\nu;\mu\nu}=\kappa^2\frac{\int d^4k}{2(2\pi)^4}
\frac{\left(k'_{\bar\mu}k_{\bar\nu}+k'_{\bar\nu}k_{\bar\mu}\right)
\left(k'_{\mu}k_{\nu}+k'_{\nu}k_{\mu}\right)}
{\left({k'}^2-m^2+i\epsilon\right)\left(k^2-m^2+i\epsilon\right)}
\label{eq2}
\end{equation},
where we set $k'=k+q$ and we take for definiteness only
fully transverse, traceless polarization states of the graviton to
be act on $\Sigma$ so that we have dropped the traces from its
vertices. Clearly, (\ref{eq2})
has degree of divergence +4. Explicit use of the Feynman rules for 
(\ref{eq1}) shows indeed that this 
superficial divergence degree
gets larger and larger as we go to higher and higher loop contributions.\par

However, there is a physical effect which
must be taken into account in a situation such as that in Fig. 1.
Specifically, the gravitational force is attractive and 
proportional to $(mass)^2$, so that
as one goes with the integration four-momentum $k$ into the 
deep Euclidean (large negative $(mass)^2$) $k^2$ regime 
( assume we have Wick rotated henceforth ),
the `attractive' force from gravity between the particle at 
a point $x$ and one at point $x'$ becomes 'repulsive' and should cause the 
respective propagator between the two points to be 
damped severely in the exact solutions of the theory. 
This suggests that we should resum the soft graviton 
corrections to the propagators in Fig. 1 to get an improved
and physically more meaningful result.\par

We point out that the procedure of resumming the propagators in a
theory to get a more meaningful and physically more accurate result
is well founded in recent years. The precision Z 
physics tests at SLC/LEP1 and the precision WW pair 
production tests at LEP2~\cite{lewwg} of the
SM are all based on comparisons between experiment and theory 
in which Dyson resummed Z and W boson propagators are essential.
For example, in Refs.~\cite{bph}, 
one can see that SM one-loop
corrections involving the exchange of Z's and W's in the respective
loops are calculated sytematically using Dyson resummed propagators
and these one-loop corrections have been found to be in
agreement with experiment~\cite{lewwg}.
Here, we seek to follow the analogous path, where instead of
making a Dyson resummation of our propagators we will make
a soft graviton resummation of these propagators.\par

We use the formulas of Yennie, Frautschi and Suura (YFS)~\cite{yfs},
which we have used with success in many higher order resummation
applications for precision Standard Model~\cite{sm} EW tests at LEP1/SLC and at LEP2\footnote{ For example, the total precision tag for the
prediction for the LEP1 luminosity small angle Bhabha scattering process
from BHLUMI4.04~\cite{BHLUMI}, which realizes ${\cal O}(\alpha^3)$LL YFS exponentiation, is $0.061\%$($0.054\%$) according as one does not ( does ) implement
the soft pairs correction as in Refs.~\cite{bhl2.30,oreste}. This
theory uncertainty enters directly in the EW precision observables
such as the peak Z cross section~\cite{lewwg} and the success of
the Z physics precision SM tests with such observables 
gives experimental validity to the YFS approach to resummation. },
to resum the the propagators in Fig. 1 and use their resummed
versions in the loop
expansion just as the Dyson resummed W and Z propagators are
used in the loop calculations in Refs.~\cite{bph}.
To this end, we need to determine the analogue for gravitons of the 
YFS virtual infrared (IR) function 
$\alpha B''\equiv\alpha B''_\gamma $ in eq.(5.16) of Ref.~\cite{yfs}
which gives the respective YFS resummation of the soft photon corrections
for the electron proper self-energy $\Sigma_F$ in QED:
\begin{equation}
\Sigma_F(p)=e^{\alpha B''_\gamma}\left[{\Sigma'}_F(p)-S^{-1}_F(p)\right]+
S^{-1}_F(p),
\end{equation}
as this latter equation implies the YFS resummed electron propagator result
\begin{equation}
iS'_F(p) = \frac{ie^{-\alpha B''_\gamma}}{S^{-1}_F(p)-{\Sigma'}_F(p)}.
\label{yfsa}
\end{equation}
Here, ${\Sigma'}_F(p)$ is proper self-energy hard photon residual
corresponding directly to the sum over the hard photon residuals
on the RHS of eq.(2.3) of Ref.~\cite{yfs}:
\begin{equation}
{\Sigma'}_F(p)=\sum_{n=1}^{\infty}{\Sigma'}_{Fn},
\end{equation}
where ${\Sigma'}_{Fn}$ is the n-loop YFS IR subtracted residual contribution
to $\Sigma_F$ which is free of virtual IR singularities and which
is defined in Ref.~\cite{yfs}, i.e., we follow directly 
the development in Ref.~\cite{yfs} as it is applied there to
the proper one-particle irreducible two-point
function for the electron, the inverse propagator function.
We stress that (\ref{yfsa}) is an {\it exact} re-arrangement of the respective
Feynman-Schwinger-Tomonaga series valid at {\it all} energies.
We also stress that, as the probability to radiate soft quanta in
high energy annihilation processes for a particle of mass m and charge e
is proportional to~\cite{yfs} $2\frac{\alpha}{\pi}\left(\ln s/m^2 - 1\right)$,
where $s=4E^2$ when $E$ is the cms energy, the effects of such resummation
become more and more important at higher and higher energies. 
From eq.(5.13) of Ref.~\cite{yfs},
we have the representation 
\begin{equation}
 \alpha B''_\gamma = \int d^4\ell\frac{S''(k,k,\ell)}{\ell^2-\lambda^2+i\epsilon}
\label{virt1}
\end{equation}
where $\lambda$ is the usual infrared regulator and the YFS virtual IR emission function is given by
\begin{equation}
S''(k,k,\ell) = \frac{-i8\alpha}{(2\pi)^3}\frac{kk'}{(\ell^2-2\ell k+\Delta+i\epsilon)(\ell^2-2\ell k'+\Delta'+i\epsilon)}{\Big|}_{k=k'}
\end{equation}
where we define $\Delta =k^2 - m^2$,~$\Delta' ={k'}^2 - m^2$. 
We see that we may also write (\ref{virt1}) as
\begin{equation}
\alpha B''_\gamma = \int \frac{d^4\ell}{(2\pi)^4}\frac{-i\eta^{\mu\nu}}{(\ell^2-\lambda^2+i\epsilon)}\frac{-ie(2ik_\mu)}{(\ell^2-2\ell k+\Delta+i\epsilon)}\frac{-ie(2ik'_\nu)}{(\ell^2-2\ell k'+\Delta'+i\epsilon)}{\Big|}_{k=k'}.
\label{virt2}
\end{equation}
Using the results
in Appendix A of Ref.~\cite{yfs}, this allows us to
write down the corresponding result
for the soft graviton virtual IR emission process, where, following 
Weinberg in Ref.~\cite{wein1} and using the Feynman rules for
(\ref{eq1}), we identify the conserved charges
in the graviton case as $e \rightarrow \kappa k_\rho$
for soft emission from k so that for the analogue of the virtual
YFS function $\alpha B''_\gamma$ we get here the graviton virtual IR
function $-B''_g(k)$ given by replacing the photon propagator in (\ref{virt2})
by the graviton propagator, $$\frac{i\frac{1}{2}(\eta^{\mu\nu}\eta^{\bar\mu\bar\nu}+
\eta^{\mu\bar\nu}\eta^{\bar\mu\nu}-\eta^{\mu\bar\mu}\eta^{\nu\bar\nu})}{\ell^2-\lambda^2+i\epsilon}$$, and by replacing the QED charges by the corresponding
gravity charges $\kappa k_{\bar\mu},~\kappa k'_{\bar\nu}$. 
In this way we get the result
\begin{equation} 
B''_g(k)= -2i\kappa^2k^4\frac{\int d^4\ell}{16\pi^4}\frac{1}{\ell^2-\lambda^2+i\epsilon}\frac{1}{(\ell^2+2\ell k+\Delta +i\epsilon)^2}
\label{yfs1} 
\end{equation}
and the corresponding scalar version of (\ref{yfsa})
as 
\begin{equation}
i\Delta'_F(k)|_{YFS-resummed} =  \frac{ie^{B''_g(k)}}{(k^2-m^2-\Sigma'_s+i\epsilon)}
\label{resum}
\end{equation}  
where the YFS soft graviton infrared subtracted residual $\Sigma'_s$ 
is in quantum gravity the scalar analogue
of the QED electron proper self-energy soft photon IR 
subtracted residual $\Sigma'_F$. As $\Sigma'_s$
starts in ${\cal O}(\kappa^2)$, we may drop it in using the result
(\ref{resum}) in calculating the one-loop effects in Fig. 1.
We also stress that, unlike its QED counterpart, this
quantum gravity soft graviton YFS IR subtracted residual is not completely
free of virtual IR singularities: the genuine non-Abelian 
soft graviton IR singularities are still present in it.
These can be computed and handled order by order in $\kappa$
in complete analogy with what is done in QCD~\cite{qcd} perturbation theory.
For the deep Euclidean regime relevant to (\ref{eq2}), we get
\begin{equation}
B''_g(k) = \frac{\kappa^2|k^2|}{8\pi^2}\ln\left(\frac{m^2}{m^2+|k^2|}\right)
\label{deep}
\end{equation}
so that, as we expected, the soft graviton resummation following
the rigorous YFS prescription causes the propagators
in (\ref{eq2}) to be damped faster than any power of $|k^2|$!
( For the massless case where the renormalized mass
$m$ vanishes, the result (\ref{deep}) is best computed 
using the customary $-\mu^2$ normalization point for
massless particles. In this way, we get $B''_g(k)=\frac{\kappa^2|k^2|}{8\pi^2}
\ln\left(\frac{\mu^2}{|k^2|}\right)$ in the deep Euclidean regime
for the massless case, where $-\mu^2$ 
can be identified as the respective (re)normalization point.
This again falls faster than any power of $|k^2|$!)
When we introduce the result (\ref{deep}) into (\ref{eq2}),
we get ( here, $k\rightarrow (ik^0,\vec k)$ by Wick rotation )
\begin{equation}
i\Sigma(q)^{1a}_{\bar\mu\bar\nu;\mu\nu}=i\kappa^2\frac{\int d^4k}{2(2\pi)^4}
\frac{\left(k'_{\bar\mu}k_{\bar\nu}+k'_{\bar\nu}k_{\bar\mu}\right)e^{\frac{\kappa^2|{k'}^2|}{8\pi^2}\ln\left(\frac{m^2}{m^2+|{k'}^2|}\right)}
\left(k'_{\mu}k_{\nu}+k'_{\nu}k_{\mu}\right)e^{\frac{\kappa^2|k^2|}{8\pi^2}\ln\left(\frac{m^2}{m^2+|k^2|}\right)}}
{\left({k'}^2-m^2+i\epsilon\right)\left(k^2-m^2+i\epsilon\right)}.
\label{eq2p}
\end{equation}
Evidently, this integral converges; so does that for Fig.1b when
we use the improved resummed propagators. This means that we
have a rigorous quantum loop correction to Newton's law
from Fig.1 which is finite and well defined.\par

More precisely, using standard resummation algebra already well-tested
in precision EW physics as cited above, we replace the
naive free propagator 
$$i\Delta_F(k)=\frac{i}{(k^2-m^2+i\epsilon)}$$ 
with the resummed ``improved Born'' propagator
\begin{equation}
\begin{split}
i{\Delta}^{(0)}_F(k)|_{YFS-resummed} &=  \frac{ie^{B''_g(k)}}{(k^2-m^2+i\epsilon)}\\
                              &=i\Delta'_F(k)|_{YFS-resummed,\Sigma'_s=0 }
\end{split}
\label{born1}
\end{equation}
{\it everywhere} in the loop expansions of our theory, with due attention
to avoid double counting, as usual. In this way,
one sees that ${\Delta}^{(0)}_F(k)|_{YFS-resummed}$ is also used in computing
the residual $\Sigma'_s(k)$ as well, so that $\Sigma'_s(k)$
is now also finite\footnote{ Note that this implies the use of the
analog of ${\Delta}^{(0)}_F(k)|_{YFS-resummed}$ for the 
virtual graviton propagators in the loops in $\Sigma'_s(k)$ as well.
The important point is that exponential factor $e^{B''_g(k)}$ is
spin independent and suppresses the corresponding resummed propagators for all
point particles.}
and indeed makes at most a small genuine two-loop,
suppressed by ${\cal O}(\kappa^2)$,
correction to the improved calculation of the one-loop corrections
in Fig. 1 that we get by using our improved propagator 
$i{\Delta}^{(0)}_F(k)|_{YFS-resummed}$ therein.
This rigorously justifies neglecting $\Sigma'_s(k)$ 
when we work to leading order
$\kappa^2$.\par

Let us examine the entire theory from (\ref{eq1}) to all orders in
$\kappa$: we write it as 
\begin{equation}
{\cal L}(x) = {\cal L}_0(x)+\sum_{n=1}^{\infty}\kappa^n{\cal L}_I^{(n)}(x)
\label{lgn1}
\end{equation}
in an obvious notation in which the first term is the free 
Lagrangian, including the free part of the gauge-fixing and ghost
Lagrangians and the interactions, including the ghost interactions, 
are the terms of ${\cal O}(\kappa^n), n\ge 1$.

Each ${\cal L}_I^{(n)}$ is itself a finite sum of terms:
\begin{equation}
{\cal L}_I^{(n)}(x)=\sum_{\ell=1}^{m_n}{\cal L}_{I,\ell}^{(n)}(x).
\label{lgn2}
\end{equation}
${\cal L}_{I,\ell}^{(n)}$ has dimension $d_{n,\ell}$.
Let $d^M_n=\max_\ell\{d_{n,\ell}\}$. As we have at least three fields
at each vertex, the maximum power of momentum at any vertex in
${\cal L}_I^{(n)}$ is $\bar{d}^M_n=\min\{d^M_n-3,2\}$ and is finite
( here, we use the fact that the Riemann tensor is only second order
in derivatives ). We will use this fact shortly. 

First we stress that, in any gauge,
if $P_{\alpha_1\cdots;\alpha'_1\cdots}$ is the respective
propagator polarization sum for a spinning particle, then the spin 
independence of the soft graviton YFS resummation exponential factor
in (\ref{born1}) yields the respective YFS resummed improved Born
propagator as
\begin{equation}
i{\cal D}^{(0)}_{F\alpha_1\cdots;\alpha'_1\cdots}(k)|_{YFS-resummed}=\frac{iP_{\alpha_1\cdots;\alpha'_1\cdots}e^{B''_g(k)}}{(k^2-m^2+i\epsilon)},
\label{spnprp}
\end{equation}
so that it is also exponentially damped at high energy in the 
deep Euclidean regime (DER). We will use this shortly as well.


Now consider any one particle irreducible vertex $\Gamma_N$ with
$[N]\equiv n_1+n_2$ amputated external legs, where we use the notation
$N=(n_1,n_2)$,
when $n_1 (n_2)$ is the respective number of graviton(scalar) external lines.
We always assume we have Wick rotated.
At its zero-loop order, 
there are only tree contributions which are manifestly UV finite.
Consider the first loop (${\cal O}(\kappa^2)$) corrections to  $\Gamma_N$.
There must be at least one improved exponentially damped propagator
in the respective loop contribution and at most two vertices
so that the maximum power of momentum in the numerator of the
loop due to the vertices is $\max\{2\bar{d}^M_1,\bar{d}^M_2\}$  and is finite. 
The exponentially damped propagator then renders the loop integrals finite 
and as there are only a finite number
of them, the entire one-loop (${\cal O}(\kappa^2)$) contribution is finite.\par

As a corollary, if $\Gamma_N$ vanishes in tree approximation, we can
conclude that its first non-trivial contributions at one-loop are
all finite, as in each such loop the exponentially damped 
propagator which must be present
is sufficient to damp the respective finite order polynomial
in loop momentum that occurs from its vertices by our arguments
above into a convergent integral.\par

\begin{figure}
\begin{center}
\epsfig{file=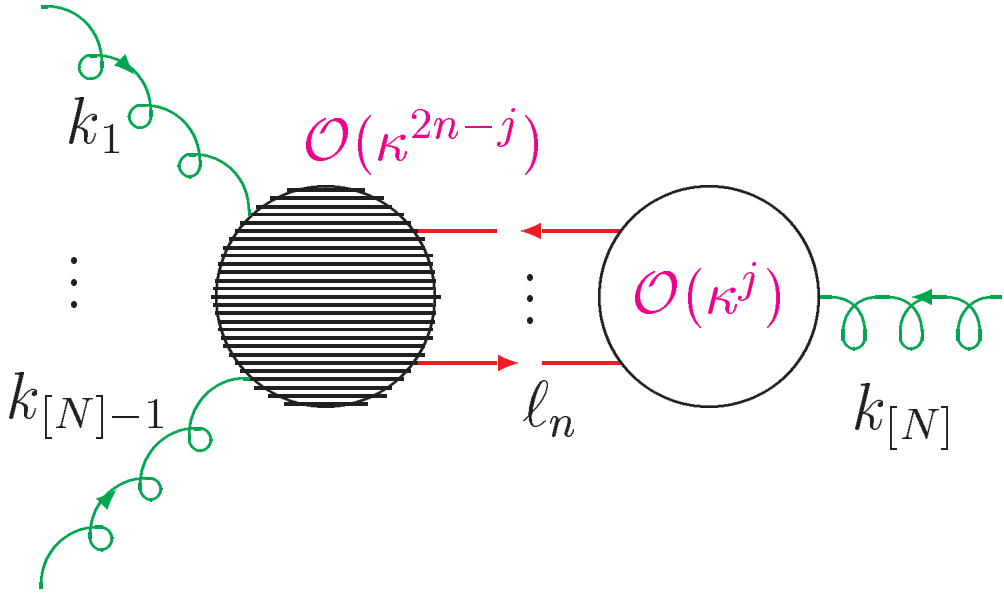,width=140mm}
\end{center}
\caption{\baselineskip=7mm     The typical contribution
we encounter in $\Gamma_N$ at the n-loop level; $\ell_n$ is the n-th
loop momentum and is precisely the momentum of the
indicated YFS-resummed improved Born propagator.
}
\label{fig2}
\end{figure}

As an induction hypothesis suppose all contributions to all $\{\Gamma_N\}$
for m-loop corrections (${\cal O}(\kappa^{2m})), m<n,$ are finite. 
At the n-loop (${\cal O}(\kappa^{2n})$) level, when the exponentially damped
improved Born propagators are taken into account, 
we argue that respective n-loop integrals are finite as follows.
First, by momentum conservation, if $\{\ell_1,\cdots,\ell_n\}$
are the respective Euclidean loop momenta, we may without loss of content
assume that $\ell_n$ is precisely the momentum of one of the
exponentially damped
improved Born propagators. The $n-1$ loop integrations over
the remaining loop variables $\{\ell_1,\cdots,\ell_{n-1}\}$
for fixed $\ell_n$ then produces the contribution of a subgraph
which if it is 1PI is a part of $\Gamma_{N+2}$ and which if it is not 1PI
is a product of the 
contributions to the respective $\{\Gamma_J\}$ and the respective 
improved YFS resummed Born propagator functions.
This is then finite by the induction hypothesis.
Here, $N+2=(n_1+2,n_2)\left((n_1,n_2+2)\right)$ according as the propagator
with momentum $\ell_n$ which we fix as multiplying the remaining subgraph
is a graviton(scalar) propagator, respectively.
The application of standard 
arguments~\cite{Royden} from Lebesgue integration
theory ( specifically, for any two measurable functions $f,g$,
$f\le g$ almost everywhere implies that $\int f \le \int g$ ) 
in conjunction with
Weinberg's theorem~\cite{wein2} guarantees that
this finite result behaves at most as
a finite power of $|\ell_n|$ modulo Weinberg's logarithms 
for $|\ell_n|\rightarrow \infty$. It follows that the remaining
integration over $\ell_n$ is damped into convergence by the
already identified exponentially damped propagator with momentum $\ell_n$.
Thus, each $n$-loop contribution to $\Gamma_N$ is finite,
from which it follows that $\Gamma_N$ is finite at $n$-loop level.
Pictorially, we illsutrate the type of situations we have
in Fig.~\ref{fig2}.
\par

We conclude by induction that all $\{\Gamma_N\}$ in our theory are finite
to all orders in the loop expansion. Of course, the sum of the respective
series in $\kappa$ may very well not actually converge but this issue
is beyond the scope of our work.\par
 
The key in the argument is the ability to isolate, for example 
in the scalar propagator, the YFS soft graviton infrared 
subtracted residual $\Sigma_s'$
order by order in our new improved loop expansion. We stress that
since we have an identity in (\ref{resum}) we have not altered
the Feynman-Schwinger-Tomonaga series; we have just re-arranged it
so that we treat $\Sigma_s'$ perturbatively relative to our
improved Born YFS-resummed propagator in complete analogy with the
treatment of the usual $\Sigma_s$ perturbatively relative to the
naive Feynman propagator $i\Delta_F$. 
To see that this makes sense,
let us write down $\Sigma_s'(k)$ explicitly in one loop order.\par

We have from (\ref{resum}) the one-loop result
\begin{equation}
\Sigma_s^{'(1)}(k)=\Sigma_s^{(1)}(k)-B''_g(k)\Delta_F^{-1}(k)
\end{equation}
where $A^{(n)}$ is the n-loop contribution to $A$. 
Using the standard methods, we see that this allows us to get
the following representation of 
$\Sigma_s^{'(1)}(k)$ in our YFS-resummed improved perturbation theory:
\begin{equation}
\begin{split}
\Sigma_s^{'(1)}(k)&= -\kappa^2\frac{\int d^4\ell}{(2\pi)^4}\Bigg{\{}\Big{[}
(2k^{\mu}k^\nu){\cal P}_{\mu\nu;\mu'\nu'}(\ell)(2k^{\mu'}k^{\nu'})\frac{\ell^2+2\ell k+2(k^2-m^2)}{\ell^2+2\ell k+k^2-m^2+i\epsilon}\\
&+\quad \Delta V^{\mu\nu}(k,\ell){\cal P}_{\mu\nu;\mu'\nu'}(\ell)(2k^{\mu'}k^{\nu'})+(2k^{\mu}k^\nu)
{\cal P}_{\mu\nu;\mu'\nu'}(\ell)\Delta V^{\mu'\nu'}(k,\ell)\\
&+\quad \Delta V^{\mu\nu}(k,\ell){\cal P}_{\mu\nu;\mu'\nu'}(\ell)\Delta V^{\mu'\nu'}(k,\ell)\Big{]}
\frac{e^{\frac{\kappa^2|(k+\ell)^2|}{8\pi^2}\ln\left(\frac{m^2}{m^2+|(k+\ell)^2|}\right)}}
{(k+\ell)^2-m^2+i\epsilon}
\frac{e^{\frac{\kappa^2|\ell^2|}{8\pi^2}\ln\left(\frac{\mu^2}{|\ell^2|}\right)}}
{\ell^2+i\epsilon}\\
&+\quad \Big{[}\frac{1}{2}(k^2-m^2)\left({{\cal P}_{\lambda\rho;}}^{\rho\lambda}(\ell)+{{\cal P}_{\lambda\rho;}}^{\lambda\rho}(\ell)-{{{{\cal P}_\lambda}^\lambda}_{;\lambda'}}^{\lambda'}(\ell)\right)\\ 
&-\quad (2k^\mu k^\nu)\left({{{\cal P}_{\mu\rho;}}^{\rho}}_\nu(\ell)+{{\cal P}_{\mu\rho;\nu}}^{\rho}(\ell)-{{\cal P}_{\mu\nu;\rho}}^\rho(\ell)\right)\Big{]}\frac{e^{\frac{\kappa^2|\ell^2|}{8\pi^2}\ln\left(\frac{\mu^2}{|\ell^2|}\right)}}
{\ell^2+i\epsilon}\Bigg{\}},
\end{split}
\label{egsg'1}
\end{equation}
where we have defined $\Delta V^{\mu\nu}(k,\ell)= k^\mu\ell^\nu + k^\nu\ell^\mu - (k^2-m^2+k\ell)\eta^{\mu\nu}$.
We see explicitly that the exponentially damped propagators render 
$\Sigma_s^{'(1)}(k)$ ultra-violet (UV) finite. 
It is clear from (\ref{egsg'1}) that $\Sigma_s'$ is indeed a small perturbation
on our results in this paper and we can neglect it in what follows.
It will be presented in detail elsewhere.
\par

Continuing to work in the transverse, traceless space for $\Sigma$,
we get, to leading order, that the graviton propagator denominator
becomes
\begin{equation}
q^2 +\frac{1}{2}q^4\Sigma^{T''}(0)+i\epsilon
\label{prop}
\end{equation}
where the self-energy function $\Sigma^T(q^2)$ follows from Fig. 1
by the standard methods ( the new type of integral
we do by steepest descent considerations for this paper )
and for its second derivative at $q^2=0$
we have the result
\begin{equation}
\begin{split}
\Sigma^{T''}(0)&= -\frac{\kappa^4m^2f_{sd}}{8\pi^3\sqrt{\pi}}\frac{240}{786432}\ln^3 x_0
              \left(1-\frac{135}{4096c_{sd}}\ln x_0\right)\\
              &\simeq -\frac{7.44m^2f_{sd}}{\pi^2 M_{Pl}^4},
\end{split}
\label{sigma}
\end{equation}
where $M_{Pl}=1.22\times 10^{19}$GeV is the Planck mass
, $x_0=\frac{\pi}{2}\frac{M^2_{Pl}}{m^2c_{sd}}$ and we work in the
leading log approximation for the big log $L=\ln x_0$.
The steepest descent factors $c_{sd},f_{sd}$ turn out to be 
$c_{sd}\simeq 12.9,~f_{sd}\simeq 5.79\times 10^2$.\par

To see the effect on Newton's potential, we Fourier transform the
inverse of (\ref{prop}) and find the potential
\begin{equation}
\Phi_{Newton}(r)= -\frac{G_N M_1M_2}{r}(1-e^{-ra})
\end{equation}
where $a=1/\sqrt{-\frac{1}{2}\Sigma^{T''}(0)}\simeq 82.6M_{Pl}$
in an obvious notation.
With the current experimental accuracy~\cite{pdg} on $G_N$ of
$\pm 0.15\%$, we see that, to be sensitive to this quantum effect, we must
reach distances $\sim 6.45\times 10^{-21}GeV^{-1}\simeq 0.13\times 10^{-33}cm$.
Presumably, in the early universe studies~\cite{kolb}, this is available
\footnote{Indeed, during the late 1960's and early 1970's, 
there was a very successful approach to
the strong interaction based on the old string theory~\cite{sch}, 
with a Regge slope
characteristic of objects of the size of hadrons, $\sim 1fm$. Experiments
deep inside the proton~\cite{taylor} showed phenomena~\cite{bj} which revealed
that the old string theory was a phenomenological model for
a more fundamental theory, QCD~\cite{gwp}. Similarly, our work suggests that
the current superstring theories~\cite{gsw,jp} of quarks and 
leptons as extended objects
of the size $\sim \ell_{M_{Pl}}=1/M_{Pl}$ may very well be also 
phenomenological models for a more fundamental theory, 
TUT (The Ultimate Theory), 
which would be revealed by
'experiments' deep inside quarks and leptons at scales well below
$\ell_{M_{Pl}}$. It is just very interesting that the early universe
may provide access to the attendant `experimental data' and the
methods we present above allow us to analyze that data, apparently.}.
This opens the distinct possibility that physics below the Planck
scale is accessible to point particle quantum field theoretic methods.
\par
    
We believe then that we have found a systematic approach to computing
quantum gravity as formulated by Feynman in Ref.~\cite{f1,f2}.
The law of Newton may now be studied on equal footing with the other
known forces in the Standard Model~\cite{sm} of $SU_c(3)\times SU_{2L}\times U_1$ interactions.\par

\section*{Acknowledgements}

We thank Profs. S. Bethke and L. Stodolsky for the support and kind
hospitality of the MPI, Munich, while a part of this work was
completed. We thank Prof. C. Prescott for
the kind hospitality of SLAC Group A as well during the beginning
stages of this work and we thank Prof. S. Jadach for useful discussions.

\section*{Notes Added}
\noindent
1. The limit of quantum gravity studied here is that of the deep Euclidean regime, i.e., the short-distance limit of the theory. This should be contrasted with the long-distance limit studied by effective Lagrangian methods such as those used in Ref.~\cite{donoghue}.\par
\noindent
2. That we get a cut-off at $\sim M_{Pl}$ in our theory means that, 
if the improved propagators we find are introduced into the 
renormalizable SM Feynman rules, the net effect will be to cut-off the 
respective Feynman integrals at  $\sim M_{Pl}$ and ,hence,
by the standard results of renormalizable quantum field theory, 
there will no effect on any of the phenomenological predictions of the 
SM at scales well below $M_{Pl}$.\par
\noindent 
3. For nonrenormalizable theories,
our methods appear to offer a `new lease' on life for many of them.
We caution that, while our improved propagators render many of them finite,
they may have other problems.

\end{document}